\documentclass[twocolumn,english,prl,showpacs]{revtex4}
\usepackage{graphicx}
\usepackage{amsmath}
\usepackage{amsfonts}
\usepackage{amssymb}
\usepackage{mathrsfs}
\usepackage{color}

\begin{document}
\title{BCS-BEC crossover in a strongly correlated Fermi gas}
\author{S. G. Bhongale,$^{12}$ M. R. Goosen,$^3$ and S. J. J. M. F. Kokkelmans$^3$}
\affiliation{$^1$Department of Physics and Astronomy MS-61, Rice University, Houston, TX 77005, USA\\
$^2$Department of Physics and Astronomy, University of New Mexico, Albuquerque, NM 87131, USA\\
$^3$Eindhoven University of Technology, P.O. Box 513, 5600MB Eindhoven, The Netherlands}

\begin{abstract}
We study the BCS-BEC crossover in the strongly correlated regime of an ultra-cold rotating two
component Fermi gas. Strong correlations are shown to generate an additional
long-range interaction which results in a modified crossover region compared to the
non-rotating situation. The two-particle correlation function
reveals a smooth crossover between the $s$-wave paired
fermionic fractional quantum Hall state and the bosonic Laughlin state.
\end{abstract}
\date{\today}
\maketitle
In recent years techniques based on Feshbach scattering resonances~\cite{fesh2}
have enabled the study of pair condensation in ultra-cold Fermi gases~\cite{fermicond1,
fermicond2,fermicond3}. For condensation to occur,
one can distinguish two distinct physical mechanisms: (1) formation
of bound pairs of fermionic atoms (molecules) which are composite
bosons and hence undergo Bose-Einstein condensation (BEC), and (2)
condensation of Bardeen-Cooper-Schrieffer (BCS) pairs in analogy
with low temperature superconductivity. In separate publications
\cite{eagles,leggett} both Eagles and Leggett argued that these
scenarios were limiting cases of a more general theory, the
so-called BCS-BEC crossover.

It was only recently that this crossover phenomenon was observed in
rotating trap experiments. A vortex lattice generated in the molecular
BEC phase was observed to persist into the BCS paired phase as the
interaction is adiabatically tuned from repulsive to attractive across
the Feshbach resonance~\cite{crossvert}.  Other experiments, where the
rotation frequency was increased such that the degenerate gas enters
the 2D regime, have led to the direct image of Tkachenko
waves~\cite{cornell}. In this fast rotation regime the effects of
strong correlations, such as those responsible for the fractional
quantum Hall (FQH) effect, remain unobserved. It has been predicted
that ultra-cold atomic systems can be brought into the FQH regime by
rotating the trap at frequencies $\Omega$ close to the trapping
frequency $\omega$~\cite{wilkins,paredes}. In addition, non-rotating
2D gases have enabled the observation of the
Berezinskii-Kosterlitz-Thouless crossover in a trap ~\cite{dalibard}
and in an optical lattice~\cite{Cornell2}.

In the present letter we study the BCS-BEC crossover in the strongly
correlated regime. Usual mean field theory breaks down as the number
of fermions becomes comparable to the number of vortices, which is
essential to reach the FQH regime. We exploit the analogy with a FQH
system and utilize the Chern-Simons gauge transformation
\cite{zhang}, which enables us to study superfluidity in the strongly
correlated regime by considering pairing of correlation-free
composite fermions. These composite particles consist of a
fermion and an even number of flux quanta. The pairing of fermions in the
strongly correlated regime is then studied by standard BCS mean field
theory applied to these composite fermions. We show that the BCS-BEC
crossover in the strongly correlated regime can be considered as a
crossover between two FQH states.

We consider a two component Fermi system consisting of a balanced
mixture of fermionic atoms in different hyperfine states represented
by $\lvert\uparrow\rangle$ and $\lvert\downarrow\rangle$ confined by a 2D
rotating harmonic trap. In the rotating frame, the Hamiltonian for
this system in the FQH regime ($\Omega-\omega\rightarrow 0^-$) in
the absence of interactions is given by
\begin{equation}
H=\sum_{\sigma}\int\!\! d{\bf r} \hat{\phi}_{\sigma}^{\dagger}({\bf
r})\frac{1}{2m}\left[{\bf p}+{\bf A({\bf
r})}\right]^2\hat{\phi}_{\sigma}({\bf r}),
\end{equation}
with $m$ the mass of a fermion and ${\bf A}=m\omega
y\hat{\text{x}}-m\omega x\hat{\text{y}}$ is analogous to the vector
potential associated with the external magnetic field in the electronic FQH
effect. The operator $\hat{\phi}_{\sigma}({\bf r})$ annihilates fermionic
atoms with spin $\sigma$ at position ${\bf r}$. In order to
simplify the above Hamiltonian we choose to work in a frame where
the vector potential ${\bf A}$ is gauged out. This is done by
performing the Chern-Simons transformation by attaching gauge field
$c_\alpha({\bf r})=-\hbar/\nu\sum_{\sigma}\int d^2{\bf r}'
\epsilon_{\alpha\beta}\hat{\rho}_{\sigma}({\bf r})({\bf r}-{\bf
r}')_{\beta}/|{\bf r}-{\bf r}'|^2$ to each bare particle resulting
in
\begin{equation}
H=\sum_{\sigma}\int \!\! d{\bf r} \hat{\varphi}_{\sigma}^{\dagger}({\bf
r})\frac{1}{2m}\left[{\bf p}+{\bf A({\bf r})}+{\bf c({\bf
r})}\right]^2\hat{\varphi}_{\sigma}({\bf r}),
\end{equation}
where $\hat{\varphi}_{\sigma}$ is the annihilation operator and
$\hat{\rho}_{\sigma}$ is the density of {\em composite} fermions of spin
$\sigma$, and $\nu$ is the filling fraction, which is the ratio of
the number of atoms to rotational flux quanta. The transformation is such that
the average gauge field $\bar{\bf c}$ cancels the external field,
thus ${\bf A}({\bf r})+{\bf c}({\bf r})={\bf A}({\bf r})+\bar{\bf
c}({\bf r})+\delta{\bf c}({\bf r})=\delta{\bf c}({\bf r})$. However,
we are left with gauge field fluctuations which are caused by density
fluctuations. If we write the interaction part of the above Hamiltonian as
$H_{\text{int}}=(1/2m)\sum_{\sigma}\int d^2{\bf
r}\hat{\varphi}_{\sigma}^{\dagger}({\bf r})\left[2 {\bf p}\delta {\bf
c}+\delta{\bf c}^2\right]\hat{\varphi}_{\sigma}({\bf r})$, we see that
even in the absence of interactions between the bare particles,
the Chern-Simons transformation gives rise to two- and three-body
interactions~\cite{lucjan}.

The two-body part has been attributed to have important consequences
for the formation of pairs in the electronic FQH effect~\cite{lucjan}.
We will neglect the induced three-body interaction and consider the
induced two-body interaction $V^{\text{ind}}$, which has the form
\begin{equation}
 V^{\text{ind}}(r)=\frac{2\hbar\omega}{\nu} \ln(r/\lambda),
\label{indint}
\end{equation}
for $r<\lambda$ and which we approximate by $V^{\text{ind}}=0$ for
$r>\lambda$. Here $\lambda$ is the typical length scale associated
with density fluctuations, and we assume that the induced interaction
is washed-out beyond this range. The long-range interaction comes on
top of the short-range two-body interaction, which is dominantly
$s$-wave interaction for fermions in different spin states. We characterize
the strength of the interaction via the scattering length $a$, that
can be varied by using a Feshbach resonance. This resonant short-range
interaction remains the same in the gauge transformed composite
particle picture \cite{bhongale}. However, the additional repulsive
long-range interaction can strongly modify the resonance properties,
and will have the effect of lifting up the bound states in the
potential and change the width of the resonance, resulting in a shift
and modification of the crossover region. This can be treated
systematically within the Chern-Simons composite particle picture,
where we first need to understand the details of this composite
particle interaction potential.

\begin{figure}[t]
  \includegraphics[scale=.35]{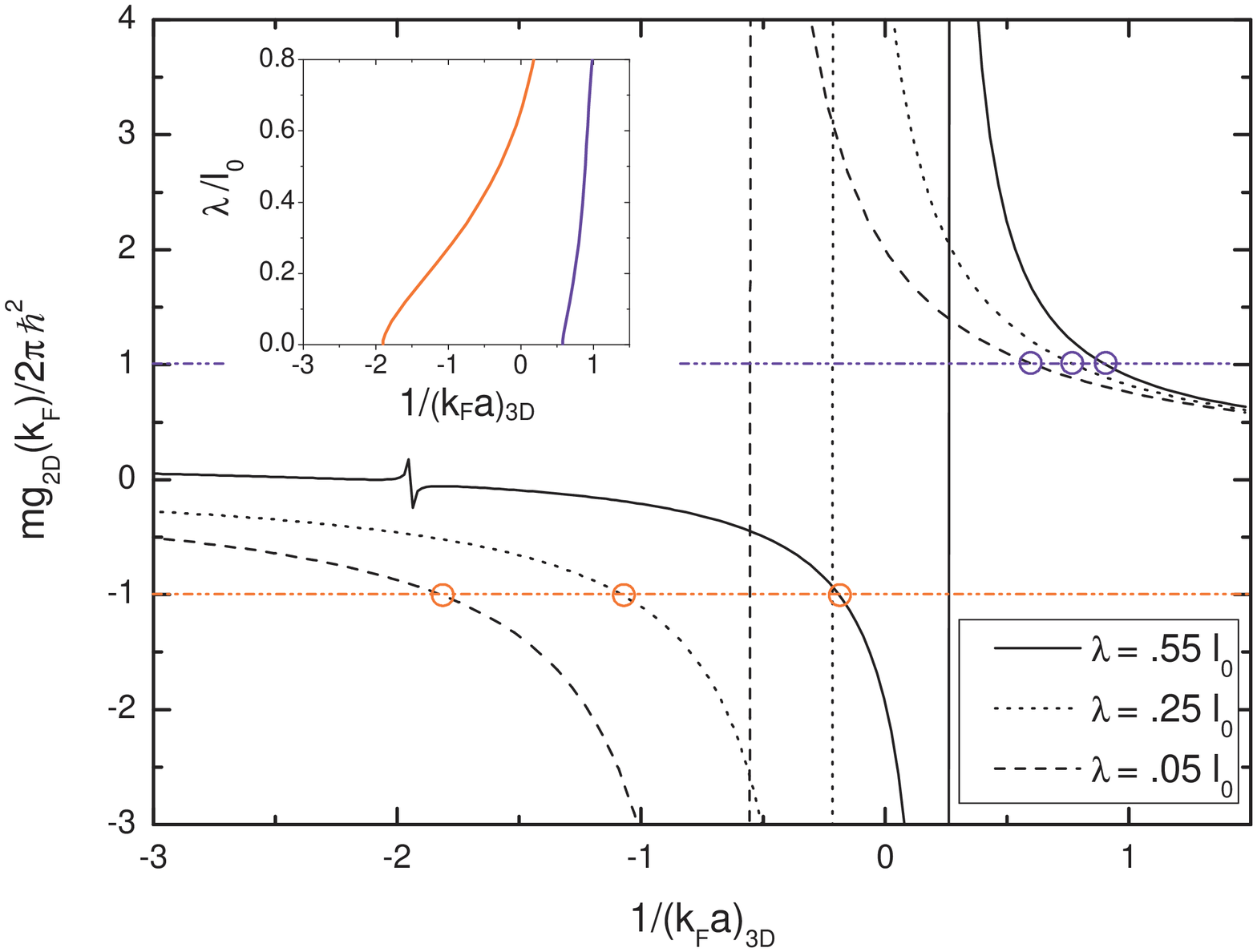}
\caption{ (Color online) The 2D interaction strength parameter
  $g_{2D}$ is plotted for different values of $\lambda$. For this
  (quasi-2D) system a confinement-induced resonance is observed for
  all values of $\lambda$. For $\lambda=.55l_0$ the atomic bound state
  is "pushed" out (for $1/(k_Fa)_{3D}\approx -1.9$) by the rotation
  induced interaction, which inevitably leads to a (narrow)
  resonance. The inset shows the shift and narrowing of the unitarity
  regime for increasing $\lambda$. The ratio of axial to radial
  trapping frequency is here set to be $10:1$. }
\label{shift}
\end{figure}

The composite fermions experience both a resonant short-range and the
logarithmic long-range interaction. We solve a 2D scattering problem
with long-range potential Eq.~(\ref{indint}), where we note that even
though the FQH effect exists in 2D systems, ultra-cold atomic systems
under extreme rotations are in fact quasi-2D. Quasi here means that
the confinement in the third dimension is strong compared to the
remaining two. Hence, the interaction at short range is 3D in nature,
and we use the relationship between the 2D and 3D scattering
length~\cite{petrov} to set a boundary condition at $r=0$. This is
done by making use of a 2D contact potential~\cite{2Dcontact}.  We
solve the 2D scattering equations as a function of relative wavenumber
$k$ and the 3D scattering length ($a_{3D}$). For every value of
$a_{3D}$ we can define an energy-dependent 2D scattering length
$a(k)$, which is related to the scattering phase shift $\delta(k)$ via
\begin{equation}
\cot \delta(k)=\frac{2}{\pi}\big(\gamma+\ln \frac{k a(k)}{2}\big),
\end{equation}
with $\gamma$ the Euler constant. We now characterize the two-body
interaction strength, which is a result of both interactions, via a
coupling parameter related to the scattering length calculated for two
particles at the Fermi energy:
\begin{equation}
g_{2D}(k_F)=\frac{2\pi\hbar^2}{m}(
\ln\frac{2}{k_Fa(k_F)}-\gamma)^{-1}.
\end{equation}
In Fig.(\ref{shift}) this coupling parameter is given as function of
the 3D scattering length, for three different values of $\lambda$. The
zero-point oscillations in the third dimension can cause density
fluctuations which result in gauge field fluctuations. We believe this
will give a natural limit to what $\lambda$ may be. On the BCS side we
see that beyond $1/(k_Fa)_{3D}= -1.9$ (for $\lambda=.55l_0$) no
two-body bound state exists. Since this implies that no many-body
pairing instability will occur \cite{randeria}, BCS pairs will be
broken.

In the following, we use a (unitary) Chern-Simons transformation to
create {\em dressed} composite fermions \cite{ezawa}. This
transformation provides a direct relation between the wavefunctions of
the fermions ($\Psi_{F}$) and composite fermions ($\Psi_{CF}$),
\begin{eqnarray}
\Psi_{F}&=&\Psi_{CF}\prod_{i<j}(z_i-z_j)^{\frac{1}{\nu}}
\prod_{i<j}(\xi_i-\xi_j)^{\frac{1}{\nu}} \prod_{i,j}(z_i-\xi_j)^{\frac{1}{\nu}}\nonumber\\
&&\times \exp\left[-\sum_k |z_k|^2/4-\sum_k
|\xi_k|^2/4\right],\label{HR}
\end{eqnarray}
where $z$ and $\xi$ are scaled in units of harmonic oscillator
length $l_0$ and represent the complex coordinate of the spin up and
spin down components respectively. The wavefunction of the dressed
composite fermions is not blurred by two-body correlations \cite{morinari}.

We will consider a system at filling fraction $\nu=1/2$ (since this
will be the experimentally most accessible regime), which means we
have transformed fermionic atoms to {\em free} interacting composite
fermions which will fill a Fermi sea. Comparison of Eq.~(\ref{HR}) to
well-known paired FQH states, e.g. Haldane-Rezayi and Moore-Read
states, shows that the pairing part of these FQH states corresponds to
the wavefunction of the composite fermions. Now in the presence of
some weakly attractive interaction, for instance caused by an
attractive atomic interaction, these composites can form BCS-like
pairs. Hence, we write the Hamiltonian for the composite fermion
system in the standard BCS form and apply (standard) mean field
theory. We start with the Bogoliubov Hamiltonian in diagonalized form
$H_{CF}=\sum_{{\bf k},\sigma}E_k\gamma_{{\bf k}\sigma}^{\dagger}
\gamma_{{\bf k}\sigma}+\mathrm{const.}$,\cite{read}, where
$\gamma^\dagger_{\sigma,{\bf k}}$ is the creation operator for an
quasi-particle with energy $E_{\bf k}=\sqrt{(\epsilon_{\bf k}-\mu)^2+
  \Delta_{\bf k}^2}$, and $\epsilon_{\bf k}$ is the single particle
kinetic energy.  Following Randeira {\em et al.}  \cite{randeria} we
solve the number and gap equation self-consistently and find the the
chemical potential equals $\mu=\epsilon_F-|E_b|/2$, where $\epsilon_F$
is the Fermi energy and $E_b$ equals the dimer binding energy in free
space inferred from the energy dependent scattering length, both at
zero temperature. In the low energy limit the gap function becomes a
constant $\Delta=\sqrt{2\epsilon_F|E_b|}$.

Since we consider $s$-wave {\em spin singlet} paired fermions, the
configuration space first quantized wavefunction for $2N=N_{\uparrow}+
N_{\downarrow}$ composite fermions can be written as $
\Psi_{CF}={\mathscr A} (\Xi_{11'}\Xi_{22'}...\Xi_{NN'})$ \cite{read},
where the anti-symmetrization is separately performed over up and down
spins (the primed and unprimed indexes) \cite{schrieffer},
$\Xi_{jj'}=\sum_{\bf k} \frac{v_{\bf k}}{u_{\bf k}} e^{i{\bf
    k}\cdot({\bf x}_j-{\bf x}_{j'})}$, and $\Big(
\begin{array}{c}
u_\mathbf{k}^2 \\
v_\mathbf{k}^2
\end{array}
\Big)=\frac{1}{2}( 1\pm(\epsilon_{\bf k}-\mu)/E_\mathbf{k})$.
Now we are able to determine the form of the pairing wavefunctions of the
(composite) fermions in the BCS and BEC regime.

To obtain a quantitative view of the crossover between the FQH states
we consider a system consisting of four fermions. On the BCS side we
use `pair' coordinates $(R_{cm},R,r)$ where $\mathbf{R}_{cm}$ denotes
the center of mass position, $\mathbf{R}$ the distance between the
centers of mass of the two pairs, and $\mathbf{r}_{jj'}={\bf x}_j-{\bf
  x}_{j'}$ the interparticle separation of the two fermions forming a
pair. These four particles have three (equivalent) ways to form pairs
which all are explicitly reproduced by Eq.~(\ref{HR}). Since the
composite fermions experience a weakly attractive interaction we know
$|E_b|\ll\epsilon_F$ and the pair wavefunction of the composite
fermions for $k_F r\gg1$ is found to be $\Xi_{jj'}\propto
\sin(k_Fr_{jj'} -\pi/4)/\sqrt{k_Fr_{jj'}}$.

Via a Feshbach resonance, we tune the $s$-wave {\em atomic} interaction such
that the composite fermions experience a weakly repulsive interaction, i.e.
$|E_b|\gg\epsilon_F$. The pair wavefunction now describes bosonic molecules
since $\Xi_{jj'}=(\Delta/2\pi\epsilon_F) K_0(\kappa r_{jj'})$ exactly equals a
(deeply bound) dimer state in 2D, where $\kappa=ik$. Advancing towards the
BEC side means the size of the molecule ($r$) has become small as compared
to the average inter-particle spacing ($\sim k_F^{-1}$) resulting in
\begin{equation}
\Psi_{F}=\Xi_{11'}\Xi_{22'}(r_{11'}r_{22'})^2 R^8 \exp\big(-R^2/4-R_{cm}^2\big).
\end{equation}
Being a degree of freedom, the coordinates $r_{11'},r_{22'}$ can be effectively
integrated out. We have obtained a state consisting of composite bosons,
which, form a BEC in the presence of the weak repulsive interaction.
Transforming to 'molecular' coordinates $Z$, which denote the positions
of the molecules in complex coordinates, results in
\begin{equation}
\Psi_{1/8}=\prod_{i<j}(Z_i-Z_j)^8
\exp\left[-\sum_k|Z_k|^2/2\right],\label{laughwave}
\end{equation}
which is a bosonic Laughlin $\nu=1/8$ state \cite{laughlin}. The exponential contains
$|Z|^2/2$ terms since the {\em molecular} harmonic oscillator length
equals $l_0/\sqrt{2}$. Now we have seen a crossover of a paired
FQH state at $\nu=1/2$ on the BCS side to a bosonic Laughlin $\nu=1/8$ state on the
BEC side for four particles. Note that on the BEC side there are half the number of particles
and the mass of each "elementary" particle (molecule) has been doubled.
This explains the change of the filling fraction, i.e. the
ratio of number of particles to the number of flux quanta (in this case
$\hbar/m$) from $\nu=1/2$ to $1/8$.

\begin{figure}[t]
  \includegraphics[scale=.35]{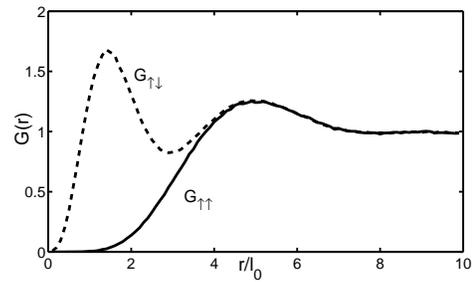}
\caption{The two particle correlation functions $G_{\uparrow\uparrow}({\bf r})$
and $G_{\uparrow\downarrow}({\bf r})$ are shown for $N_{\uparrow}=N_{\downarrow}=
100$ and $\eta=l_0$. Notice that for large $r/l_0$, $G_{\uparrow\uparrow}({\bf r})-G_{\uparrow
\downarrow}({\bf r})\rightarrow 0$.} \label{corr1}
\end{figure}

This crossover between two strongly correlated states is believed to be
valid for a larger number of fermions. To verify this, we have calculated
the two-particle correlation function $G({\bf r}_1-{\bf r}_2)=\int..\int
d^2{\bf r}_3..d^2{\bf r}_N |\Psi_{F}|^2$ for $N_{\uparrow}=N_{\downarrow}=
100$ particles using a metropolis Monte-Carlo algorithm. Since we know the
form of the paired FQH state as a function of the BCS coherence length ($\eta=
\hbar^2 k_F/\pi m\Delta$) we parameterize the Monte-Carlo calculation by
$\eta$. In Fig.~\ref{corr1}, we plot both $G_{\uparrow\uparrow}({\bf r})$
and $G_{\uparrow\downarrow}({\bf r})$ for $\eta/l_0=1$. We see that $G_{\uparrow
\downarrow}({\bf r})$ shows a peaked behavior for small $r$ that is absent in
$G_{\uparrow\uparrow}({\bf r})$. At the same time for large $r$, $G_{\uparrow
\uparrow}({\bf r})-G_{\uparrow\downarrow}({\bf r})\rightarrow 0$ implying the
existence of a sum rule special to the Haldane Rezayi like state of Eq.~(\ref{HR}),
valid throughout the region of our current interest.

\begin{figure}[t]
  \includegraphics[scale=.35]{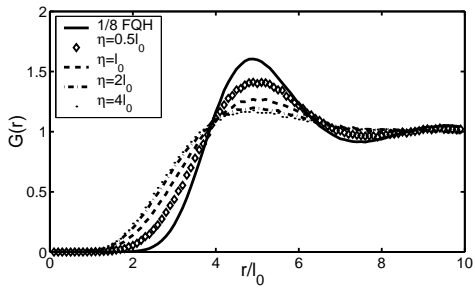}
\caption{In the strong pairing limit the correlation
function $G_{\uparrow\uparrow}({\bf r})$ has a limiting form which equals
the $G({\bf r})$ form of the $1/8$ FQH wavefunction Eq.(\ref{laughwave}).
} \label{corr2}
\end{figure}

Since in the $k\rightarrow 0$ limit, the $s$-wave $T$ matrix is a smooth function
of the $a_{3D}$ $s$-wave scattering length \cite{servaas}, the functional form of
the $T$ matrix and hence the gap $\Delta$ near the Feshbach resonance will remain
unchanged hinting a smooth crossover. The crossover behavior of the correlation
function is clear from Fig.~\ref{corr2}, which shows that as $\eta$ becomes small
compared to $l_0$, $G_{\uparrow\uparrow}({\bf r})$ gets modified continuously and
tends towards a limiting form. However the most important point to note is that
the limiting form of $G_{\uparrow\uparrow}({\bf r})$ is exactly that of the
$G({\bf r})$ for the $(1/8)$-FQH state given by the Laughlin form in Eq.~(\ref{laughwave}).

In conclusion, we have shown that the strong correlations associated
with rapid rotations can cause strong modifications to the
crossover, altering the width and position of the crossover.
Additionally we have seen the fluctuation induced interaction can
be sufficiently strong to break BCS pairs and consequently lose
superfluidity. Using $s$-wave paired FQH wavefunctions, we have shown
that the crossover is smooth and the paired FQH state of fermions smoothly
goes over to $1/8$ bosonic FQH state of molecules when one goes across
the Feshbach resonance such that $\eta\ll l_0$.

A detailed calculation of the crossover physics of this region will
require the exact nature of the rotation-induced long-range interaction Eq.~(\ref{indint}).
Within such an treatment for instance Nozi\`{e}res-Schmitt-Rink
calculations of the crossover region \cite{nsr} can be carried out.
Also these calculations can be extended to situations with $p$- and $d$- wave
pairing schemes in ultra-cold Fermi gases. These scenarios, while having close
resemblance with for example the 5/2 FQH effect, will be extremely
useful and will be dealt with in a future publication.

It is experimentally difficult to reach the FQH regime ~\cite{bloch}.
A promising technique is the combination of rotation with optical
lattices~\cite{bhat}, which is for instance able to reduce the number
of particles per vortex.

At the same time paired FQH states such as $5/2$ are known to
possess exotic non-abelian quasi-particles excitations. While
existence of non-abelian statistics is the basis for topological
scheme of implementing quantum logic in a quantum computer, the
$5/2$ state is proved to be computationally non-universal. However,
there have been proposals \cite{bravyi} in which this symptom can be
remedied by dynamically tuning-in additional non-topological
interactions. Dynamic control, while hard in the solid state
configurations of the FQH effect, transitions between different
FQH states like the one discussed here may be extremely useful
for implementing such topological schemes.

SB acknowledges financial support from the ONR, Contract
No. N00014-03-1-0508, and the W. M. Keck Program in Quantum Materials
at Rice University. SK acknowledges financial support from the NWO.

\bibliography{crossover_mg4}

\end{document}